\documentstyle[12pt]{article}
\def\({\left(}
\def\){\right)}
\def\l|{\left|}
\def\r|{\right|}
\def\<{\left<}
\def\>{\right>}
\def\[{\left[}
\def\]{\right]}
\def\ll{\left\{}
\def\lr{\right\}}

\def\qb{\bar{q}}
\def\tivj{t \rightarrow q_i \, X_{jj}}
\def\tikk{t \rightarrow q_i \, (q_k \bar{q}_k)}
\def\tbw{t \rightarrow b \, W}
\def\kkvj{(q_k \bar{q}_k) \rightarrow X_{jj}}
\def\evj{{\epsilon_{X_{jj}}}}
\def\vtj{V_{tj}}
\def\vij{V_{ij}}
\def\vtk{V_{tk}}
\def\vik{V_{ik}}
\def\vtb{V_{tb}}
\def\vib{V_{ib}}

\def\vtd{V_{td}}
\def\vid{V_{id}}
\def\acp{{\cal A}_{\rm CP}}
\def\br{{\cal B}}
\def\m{{\cal M}}
\def\t{{\cal T}}

\def\f{{\cal F}}
\def\k{{\cal K}}
\def\im{{\rm Im}}
\def\re{{\rm Re}}
\def\v{{X_{jj}}}
\def\fv{{f_{X_{jj}}}}
\def\fvh{{\hat{f}_{X_{jj}}}}
\def\mt{{m_t}}
\def\mw{{m_W}}
\def\mwh{{\hat{m}_W}}
\def\mv{{m_{X_{jj}}}}
\def\mvh{{\hat{m}_{X_{jj}}}}

\def\mb{{\hat{m}_b}}

\def\mj{{m_j}}
\def\mx{{\hat{m}_x}}
\def\my{{\hat{m}_y}}
\def\mz{{\hat{m}_z}}

\def\flxy{{g^{xy}}}
\def\gfh{{\hat{G}_F}}

\def\journal#1#2#3#4{{\it #1} {\bf #2} (#3) #4}
\def\prl{Phys. Rev. Lett.}

\def\np{Nucl. Phys.}

\def\pr{Phys. Rev.}
\def\prp{Phys. Rep.}

\title{Branching-ratio and CP violation in semi-inclusive 
	flavor-changing top decays}
\author{\vspace{5mm}\\
        {\bf L. T. Handoko}\thanks{
	E-mail address : handoko@p3ft.lipi.go.id} \\
	\vspace{2mm}\\
	Laboratory for Theoretical Physics and Mathematics \\
	Indonesian Institute of Sciences \\
	Kom. PUSPIPTEK Serpong P3FT--LIPI, Tangerang 15310, Indonesia}
\date{}

\begin{document}

\maketitle
\begin{picture}(0,0)
       \put(310,255){LFTMLIPI-139801}
       \put(310,240){January 1998}
\end{picture}

\thispagestyle{empty}

\begin{abstract}
Semi-inclusive top decays, $t \rightarrow q_i \, \v$, are 
examined in the framework of the standard model, with 
$q_i = u \; {\rm or} \; c$, $\v$ :
vector meson formed by $q_j \qb_j$ ($q_j = d, s, b$).
By using a simple model of hadronization, that is 
$\< 0 \l| \qb_j \, \gamma^\mu \, q_j \r| \v \> = \mv \, \fv \, \evj^\mu$, 
the total branching ratio is expected to be 
$\br(\tivj) < \( {\l| \vij^\ast \, \vtj \r|}/{\l| \vtb \r|}\)^2
\times O({10}^{-3})$ depends on the size of the coupling 
constant $\fv$. On the other hand, the CP asymmetry 
should be induced only by the long-distance effects of 
intermediate state $q_k \qb_k$ ($q_k = d, s, b$) that subsequent 
scatters into the vector meson. The analytical results including 
the continuum contributions are given with keeping all masses. 
However, it yields the CP asymmetry to be small, i.e. $\acp (\tivj) < 
{\im \[ \( \vij^\ast \, \vtj \)^\ast \( \vik^\ast \, \vtk \)\]}/{
\l| \vij^\ast \, \vtj \r|^2} \times O({10}^{-2})$ for an identified 
$q_k \qb_k$ intermediate state, and suppressed more again by a 
factor of $\sim {\( {m_b}^2 - {m_s}^2\)}/{\mv^2}$ for summing up all 
of them. Particularly, $\acp (t \rightarrow u \, \Upsilon) 
\simeq \pm 1\%$, induced by the intermediate state $d\bar{d}$ 
or $s\bar{s}$.
\end{abstract}

\vspace*{5mm}
PACS number(s) : 14.65.Ha

\clearpage

After the discovery of the top quark and the determination
of its mass \cite{top}, more valuable information should be
expected from some classes of both tree-level and rare top
decays. Because of its large mass scale, top quark will decay before
hadronization and also the effects of QCD should be
unambiguous and presumably very small. Then the precise
determination of the modes may provide important data with
less uncertainties.

In general, the rare decays are interesting because of
its sensitiveness to the new physics. However, especially in the
case of rare top decays, the branching ratio in the
framework of standard model (SM) is largely suppressed by
the GIM mechanism, e.g. ${\cal B}(t \rightarrow \gamma, H,
g) \sim {10}^{-7}\sim{10}^{-10}$ as pointed out in
\cite{2hdm}. Further, the class of these rare top decays are
expected to be enhanced significantly by some new contributions due to
physics beyond the SM, e.g. \cite{2hdm, mssm}. However, 
they are still anticipated to be small and only in few 
cases \cite{mssmwr} it is expected to be measureable at 
future top factories like the upgraded Tevatron and LHC 
\cite{lhc}. 

The purpose of this paper is to study the semi-inclusive top
decays, that is $\tivj$ with $q_i = u, c$ and $\v$ is any 
vector meson formed by $q_j \qb_j$ bound state as 
$\Upsilon$ for $b \bar{b}$, $\phi$ for $s \bar{s}$ and so on.
Since the processes occur in tree-level, the measurements 
can be achieved easier in future experiments. The 
interest is, especially, focused on the CP violation in the 
modes by considering CP asymmetry 
\begin{equation}
	\acp(\tivj) \equiv \frac{
	 	\( \Gamma - \bar{\Gamma} \) (\tivj)}{ 
		\( \Gamma + \bar{\Gamma} \) (\tivj)} 
		\equiv \frac{\Delta}{\Sigma} \; ,
	\label{eqn:a}
\end{equation}
with $\bar{\Gamma}$ is the complex conjugate of $\Gamma$,
while in general 
\begin{eqnarray}
	\Delta & = & -2 \, \im \( \alpha^\ast \, \beta \) \,
		\im \( {\m_1}^\ast \, \m_2 \) \; ,
	\label{eqn:delta}\\
	\Sigma & = & \l| \alpha \r|^2 \, \l| \m_1 \r|^2 
		+ \l| \beta \r|^2 \, \l| \m_2 \r|^2
		+ 2 \, \re \( \alpha^\ast \, \beta \) \, 
			\re \( {\m_1}^\ast \, \m_2 \) \; ,
	\label{eqn:sigma}
\end{eqnarray}
if one writes the amplitude as $\m \equiv \alpha \, \m_1 + \beta \, \m_2$.
Hence, the imaginary parts of $\( \alpha^\ast \, \beta \)$ 
and $\( {\m_1}^\ast \, \m_2 \)$ are required to be non-zero 
coincidently in order to have non-zero CP asymmetry. 

In the framework of the SM, the decay is realized by a 
tree $W-$exchange diagram. Then the lowest order decay
amplitude of $\tivj$ is obtained,
\begin{equation}
	\m^{(0)}(\tivj) = \sqrt{2} \, G_F \, \( \vij^\ast \, \vtj \) \, 
	\frac{\fv \, \mv \, \mw^2}{r^2 - \mw^2 + i \, \Gamma_W \, \mw} \, 
		{\evj^\ast}^\mu \,
	\[ \qb_i \, \gamma_\mu \, L \, t \] \; ,
	\label{eqn:m0}
\end{equation}
with $L = {(1 - \gamma_5)}/2$ and $r^2$ is the square of 
momentum transfer to $W-$boson. The vector meson is factorized as 
\begin{equation}
	\< 0 \l| \qb_j \, \gamma^\mu \, q_j \r| \v \> 
		= \mv \, \fv \, \evj^\mu \; ,
	\label{eqn:fv}
\end{equation}
where $\fv$ is a constant with dimension of mass
and $\evj^\mu$ denotes the polarization vector. 
Remark that the color-factor in Eq. (\ref{eqn:m0}) has
been omitted, and then should be considered as an 
unknown parameter absorbed in $\fv$.

It is straightforward to obtain the total branching 
ratio for $\tivj$ decays by assuming the charged-current
$\tbw$ to be a dominant mode in the top quark decay. 
This leads 	
\begin{equation}
	\br(\tivj) \simeq \frac{\Gamma(\tivj)}{\Gamma(\tbw)}.
	\label{eqn:br}
\end{equation}
In the top quark center-of-mass system, the denominator reads
\begin{equation}
	\Gamma(\tbw) = \frac{\gfh \, \mt}{8 \, \sqrt{2} \, \pi} \, 
		\l| \vtb \r|^2 \, \sqrt{g^{bW}} \, {\f_1}^{bW} \; ,
	\label{eqn:dwtbw}
\end{equation}
where 
\begin{eqnarray}
	\flxy & \equiv & 1 + \mx^4 + \my^4 - 
		2 \, \( \mx^2 + \my^2 + \mx^2 \, \my^2 \) \; , 
	\label{eqn:flxy}\\
	{\f_1}^{xy} & \equiv & 
	 \( 1 - \mx^2 \)^2 + \( 1 + \mx^2 \) \, \my^2 - 2 \, \my^4 \; .
	\label{eqn:f1}
\end{eqnarray}
A caret means normalization with $\mt$. 
This agrees with some literatures in $\mb \simeq 0$ limit. The value 
is, $\Gamma(\tbw) = \( 1.72 \times \l| \vtb \r|^2 \)$(GeV) for 
$\mt = 180$(GeV), $\mw = 80.33$(GeV) and $m_b = 4.3$(GeV) \cite{pdg}. 
Note that including the bottom-mass reduces the width by $\sim 2\%$.
On the other hand, the nominator is 
\begin{equation}
	\Gamma^{(0)} (\tivj) = 
		\frac{\gfh^2 \, \fvh^2 \, \mwh^4 \, \mt}{8 \, \pi} \, 
		\l| \vij^\ast \, \vtj \r|^2 \, 
		\sqrt{g^{i\v}} \, {\f_1}^{i\v} \, 
		\l| {\f_2}^{i\v W} \r|^2 \; ,
	\label{eqn:dw0}
\end{equation}
under assumptions that $E_\v = 2 \, E_j$ and $\mv = 2 \, \mj$, 
with $E$ denotes the time-component of four-momentum. Here 
${\f_2}^{i\v W}$ is the $W-$boson propagator contribution,  
\begin{equation}
	{\f_2}^{xyz} \equiv \[
	\frac{1}{2} \( 1 + \mx^2 - \my^2 \) - \mz^2 
	+ i \, {\hat{\Gamma}_z} \, \mz \]^{-1} \; . 
	\label{eqn:f2}
\end{equation}
Then, the branching ratio for $\tivj$ decay defined in 
Eq. (\ref{eqn:br}) is found to be, $\br(\tivj) \simeq 0.26 \times 
\fvh^2 \, \( {\l| \vij^\ast \, \vtj \r|}/{\l| \vtb \r|} \)^2$.
Remark that the dependences on the description of the 
flavors ($i, j$) are tiny ($O({10}^{-3})$ level). This means that
the differences between them are mostly determined by the 
CKM matrix elements. Assuming $\fvh \sim O(\mvh)$,
the branching ratio can reach $\sim  
\( {\l| \vij^\ast \, \vtj \r|}/{\l| \vtb \r|} \)^2 \times O({10}^{-3})$.

There is no CP violation in the modes as long as considering
only the lowest order diagram, because there is only one 
amplitude with single phase that governs the processes. So, 
one needs other sources with different 
phases to induce non-zero CP asymmetry. As mentioned 
before, unfortunately the loop contributions like the
penguin diagrams, are also vain because of large suppression
due to GIM mechanism. Then, one must proceed to consider the
long-distance contributions arise from any on-mass-shell 
intermediate states that then scatter into $\v$ through 1-photon 
exchange. Anyway, 1-gluon mediated scattering may be
dominate, but unfortunately the QCD scattering can occur 
at the order of ${\alpha_s}^3$ when, as the present paper, 
one considers only color singlet vector mesons as the final 
states. Of course, it may still be comparable with the 
electromagnetic scattering as occured in the 
$q \qb \rightarrow J/\psi$ \cite{qqjp}, however
since both contributions are similar and the electromagnetic
one is simpler to calculate, only 1-photon mediated
scattering are going to be discussed, i.e.
\begin{equation}
	t \longrightarrow q_i \, \( q_k \qb_k \) 
		\stackrel{\rm QED}{\longrightarrow} \v \; \; .
	\label{eqn:kkvj}
\end{equation}
Here, all intermediate states ($d \bar{d}, s \bar{s}, 
b \bar{b}$) can contribute to the processes. Since the top
quark is not expected to form any bound states and $q_i$ 
appears as a jet, the effects of the rescattering of 
final states, i.e. the intermediate states that have same quark 
content as the final states, to CP asymmetry are not important. In this 
meaning, the situation is different with the $B$ decays
\cite{fsi, bdv}. However, there will be two cases that make 
the size of CP asymmetries to be largely different. 
The first case is when one can identify and then pick up 
one of the intermediate states, and the second is when 
one can not identify and sum up all of them.
In the former case the CP asymmetry can be large,
while in the later one it will be suppressed by GIM mechanism.
Both cases will be discussed in detail later. Further, related with the 
present interest, it is sufficient enough to calculate the 
continuum contributions of these intermediate states.

Since the long-distance contributions must be smaller by a factor 
of $\alpha$, then it can be treated perturbatively.  
Hence the amplitude including one of the intermediate 
states can be expressed as 
\begin{equation}
	\m(\tivj) = \m^{(0)}(\tivj) + \frac{i}{2} \, 
		\m^{(0)}(\tikk) \, \t(\kkvj) \; .
	\label{eqn:tm}
\end{equation}
Similar with Eq. (\ref{eqn:m0}), $\m^{(0)}(\tikk)$ is
given as
\begin{equation}
	\m^{(0)}(\tikk) = \sqrt{2} \, G_F \, \( \vik^\ast \, \vtk \) \, 
	\frac{\mw^2}{r^2 - \mw^2 + i \, \Gamma_W \, \mw} \, 
		\[ \qb_k \, \gamma^\mu \, q_k \] \, 
		\[ \qb_i \, \gamma_\mu \, L \, t \] \; ,
	\label{eqn:m0p}
\end{equation}
and the 1-photon mediated transition amplitude is 
\begin{equation}
	\t(\kkvj) = \alpha \, \pi \, Q_j \, Q_k \, 
		\frac{\fv}{\mv} \, \evj^\mu \, 
		\[ \qb_k \, \gamma_\mu \, q_k \] \; .
	\label{eqn:t}
\end{equation}
It is obvious that the additional term in Eq. (\ref{eqn:tm})
induces different phase, except for $q_k = q_j$. Therefore the
interference term between them will induce CP asymmetry as expected.

First, let us give the expression for the denominator in 
Eq. (\ref{eqn:a}). Since the perturbative term in 
Eq. (\ref{eqn:tm}) is supposed to be small compared with 
$\m^{(0)}$, it can be ignored. Then the result is straightforward, 
\begin{equation}
	\Sigma \simeq \Gamma^{(0)} (\tivj) \; .
	\label{eqn:sigmar}
\end{equation}
After that, one can consider the nominator with one 
intermediate state of $q_k \qb_k$ corresponds with the first
case mentioned above. It is calculated to be
\begin{eqnarray}
	\Delta_k 
	& = & - \frac{
	Q_j \, Q_k \, \alpha \, \gfh^2 \, \fvh^2 \, \mwh^4 \, \mt}{
		\mvh^2} \, 
	\sqrt{g^{i\v}} \, \l| {\f_2}^{i\v W} \r|^2 \, 
	\nonumber \\
	& & \times 
	\im \[ \( \vij^\ast \, \vtj \)^\ast \( \vik^\ast \, \vtk \) \] \,
	\[ {\f_3}^{i\v} + {\hat{m}_k}^2 \, {\f_4}^{i\v} \] \; ,
	\label{eqn:deltak}
\end{eqnarray}
where the dependence of the intermediate state mass, $m_k$, 
has been collected in $\f_4$, that is
\begin{eqnarray}
	{\f_3}^{xy} & = & \my^2 \, \[ 
	\frac{1}{4} \, \( \flxy + 4 \, \my^2 \) \, 
		\sqrt{\flxy + 4 \, \mx^2}
	- \( 1 - \mx^2\) \, \sqrt{\flxy + 4 \, \my^2} \right.
	\nonumber \\
	& & \left. \; \; \; \; \; \; \; \; 
	- \frac{1}{2} \, \( 1 - \mx^2\)^2 + \frac{1}{2} \, \my^4 \] \; ,
	\label{eqn:f3} \\
	{\f_4}^{xy} & = & 2 \, \[ 
	\( 1 - \mx^2 - \my^2 \) \, \sqrt{\flxy + 4 \, \my^2}
	- \( 1 - \mx^2\)^2 \right.
	\nonumber \\
	& & \left. \; \; \; \; \; \; \; \; \; \; \; 
	+ 2 \, \my^2 \, \( 1 + \mx^2 - \frac{1}{2} \, \my \) \] \; .
	\label{eqn:f4}
\end{eqnarray}
On the other hand, for the second case, using the unitarity 
relation of the CKM matrix, i.e. $\sum_{k=d,s,b} \vik^\ast \, \vtk = 0$, 
and summing up all intermediate states, one finds
\begin{eqnarray}
	\Delta_{\rm tot} 
	& = & \frac{1}{3} \, \frac{
	Q_j \, \alpha \, \gfh^2 \, \fvh^2 \, \mwh^4 \, \mt}{\mvh^2} \, 
		\sqrt{g^{i\v}} \, \( {\f_2}^{i\v W} \)^2 \, {\f_4}^{i\v}
	\nonumber \\
	& & \times \ \ll 
	\im \[ \( \vij^\ast \, \vtj \)^\ast \( \vib^\ast \, \vtb \) \] \, 
		\[ {\hat{m}_b}^2 - {\hat{m}_s}^2 \] 
	\right. 
	\nonumber \\
	& & \left. \; \; \; \; \;  \; \;  
	- \im \[ \( \vij^\ast \, \vtj \)^\ast \( \vid^\ast \, \vtd \) \] \, 
		\[ {\hat{m}_s}^2 - {\hat{m}_d}^2 \] \lr \; ,
	\label{eqn:deltat}
\end{eqnarray}
since $Q_d = Q_s = Q_b = {-1}/3$ and the interference terms 
between higher-order terms are neglected. Now, combining the results
in Eqs. (\ref{eqn:sigmar})$\sim$(\ref{eqn:deltat}), 
the CP asymmetry defined in Eq. (\ref{eqn:a}) can be found to be,
\begin{eqnarray}
	\acp (\tivj) & \sim & \ll
	\begin{array}{lcl}
		-\k^{ijk} \times 0.007 & , & 
			{\rm with \; one} \; (q_k \qb_k) \\
		-\k^{ijb} \times O({10}^{-5}) & , & 
			{\rm with \; all} \; (q_k \qb_k) \\
	\end{array} 
	\right.
	\label{eqn:result}
\end{eqnarray}
where $\k^{ijk} \equiv 
{\im \[ \( \vij^\ast \, \vtj \)^\ast \( \vik^\ast \, \vtk \)\]}/{
\l| \vij^\ast \, \vtj \r|^2}$. 
Further, the results in Eqs. (\ref{eqn:dw0}), 
(\ref{eqn:f4}) and (\ref{eqn:deltat}) show that in case of summing up
all intermediate states, the CP asymmetry will be suppressed by
a factor of $\sim {\( {m_b}^2 - {m_s}^2\)}/{\mv^2}$, if we put 
${\( {m_s}^2 - {m_d}^2\)}/{\mv^2} \sim 0$.
For example, definite expressions 
for $\k^{ibk}$ and $\k^{isk}$, corresponds with the final states 
$\Upsilon$ and $\phi$, can be written as
\begin{equation}
\begin{array}{lclclc}
	\k^{ubd} = \frac{-\eta}{\rho^2 + \eta^2} & , & 
	\k^{ubs} = \frac{\eta}{\rho^2 + \eta^2} & , &
	\k^{ubb} = 0 & , \\
	\k^{cbd} = \lambda^2 \, \eta & , & 
	\k^{cbs} = 0 & , & 
	\k^{cbb} = 0 & , \\ 
	\k^{usd} = \eta & , & 
	\k^{uss} = 0 & , & 
	\k^{usb} = -\eta & , \\
	\k^{csd} = -\lambda^2 \, \eta & , & 
	\k^{css} = 0 & , & 
	\k^{csb} = 0 & ,
\end{array}
	\label{eqn:kijk} 
\end{equation}
by using the Wolfenstein parametrization. The largest factor
is one of the decay $t \rightarrow u \, \( d \bar{d}, s\bar{s} \) 
\rightarrow \Upsilon$, that is $\k^{ubd} = 1.65$ and 
$\k^{ubs} = -1.65$ for  
$(\rho, \eta) = (0.3, 0.34)$. This leads the CP asymmetry 
to be 
$ \l| \acp ( t \rightarrow u \, (d\bar{d}, s\bar{s}) \rightarrow 
\Upsilon ) \r| \simeq 1\%$. 

In conclusion, in the framework of the SM, it is shown that 
the rate of $\tivj$ may be large enough to be measured in 
the future top factories like LHC and the upgraded Tevatron,
while the CP asymmetries are too small to be detected. 
However, inversely it means that the modes are clean and good probes 
to see CP asymmetries raise in new physics beyond the SM,
for example, in the supersymmetric (SUSY) model.
The enhancement of CP asymmetry in $\tivj$ due to the 
contributions of scalar fermions in the SUSY model 
will be published in next paper \cite{handoko}.

The author thanks J. Hashida for useful discussion during the work. 
A part of this work is supported by Ministry of Education 
and Culture of Japan under Monbusho Fellowship Program.

\end{document}